\documentclass{emulateapj}

\shorttitle{CANGAROO-III observations of HESS J1804$-$216}
\shortauthors{Higashi et al.}

\begin{document}
\title{Observation of Very High Energy gamma rays from HESS J1804$-$216 with CANGAROO-III Telescopes}
\author{
Y.~Higashi,\altaffilmark{1}
H.~Kubo,\altaffilmark{1}
T.~Yoshida,\altaffilmark{2}
R.~Enomoto,\altaffilmark{3}
T.~Tanimori,\altaffilmark{1}
P.~G.~Edwards,\altaffilmark{6}
T.~Naito,\altaffilmark{9}
G.~V.~Bicknell,\altaffilmark{4}
R.~W.~Clay,\altaffilmark{5}
S.~Gunji,\altaffilmark{7}
S.~Hara,\altaffilmark{8}
T.~Hara,\altaffilmark{9}
T.~Hattori,\altaffilmark{10}
S.~Hayashi,\altaffilmark{11}
Y.~Hirai,\altaffilmark{2}
K.~Inoue,\altaffilmark{7}
S.~Kabuki,\altaffilmark{1}
F.~Kajino,\altaffilmark{11}
H.~Katagiri,\altaffilmark{12}
A.~Kawachi,\altaffilmark{10}
T.~Kifune,\altaffilmark{3}
R.~Kiuchi,\altaffilmark{3}
J.~Kushida,\altaffilmark{10}
Y.~Matsubara,\altaffilmark{13}
T.~Mizukami,\altaffilmark{1}
Y.~Mizumoto,\altaffilmark{14}
R.~Mizuniwa,\altaffilmark{10}
M.~Mori,\altaffilmark{3}
H.~Muraishi,\altaffilmark{15}
Y.~Muraki,\altaffilmark{13}
T.~Nakamori,\altaffilmark{1}
S.~Nakano,\altaffilmark{1}
D.~Nishida,\altaffilmark{1}
K.~Nishijima,\altaffilmark{10}
M.~Ohishi,\altaffilmark{3}
Y.~Sakamoto,\altaffilmark{10}
A.~Seki,\altaffilmark{10}
V.~Stamatescu,\altaffilmark{5}
T.~Suzuki,\altaffilmark{2}
D.~L.~Swaby,\altaffilmark{5}
G.~Thornton,\altaffilmark{5}
F.~Tokanai,\altaffilmark{7}
K.~Tsuchiya,\altaffilmark{1}
S.~Watanabe,\altaffilmark{1}
Y.~Yamada,\altaffilmark{11}
E.~Yamazaki,\altaffilmark{10}
S.~Yanagita,\altaffilmark{2}
T.~Yoshikoshi,\altaffilmark{3} and
Y.~Yukawa\altaffilmark{3}
}
\email{higashi@cr.scphys.kyoto-u.ac.jp}

\altaffiltext{1}{ Department of Physics, Kyoto University, Sakyo-ku, Kyoto 606-8502, Japan} 
\altaffiltext{2}{ Faculty of Science, Ibaraki University, Mito, Ibaraki 310-8512, Japan} 
\altaffiltext{3}{ Institute for Cosmic Ray Research, University of Tokyo, Kashiwa, Chiba 277-8582, Japan} 
\altaffiltext{4}{ Research School of Astronomy and Astrophysics, Australian National University, ACT 2611, Australia} 
\altaffiltext{5}{ School of Chemistry and Physics, University of Adelaide, SA 5005, Australia} 
\altaffiltext{6}{ CSIRO Australia Telescope National Facility, Narrabri, NSW 2390, Australia} 
\altaffiltext{7}{ Department of Physics, Yamagata University, Yamagata, Yamagata 990-8560, Japan} 
\altaffiltext{8}{ Ibaraki Prefectural University of Health Sciences, Ami, Ibaraki 300-0394, Japan} 
\altaffiltext{9}{ Faculty of Management Information, Yamanashi Gakuin University, Kofu, Yamanashi 400-8575, Japan} 
\altaffiltext{10}{ Department of Physics, Tokai University, Hiratsuka, Kanagawa 259-1292, Japan} 
\altaffiltext{11}{ Department of Physics, Konan University, Kobe, Hyogo 658-8501, Japan} 
\altaffiltext{12}{ Department of Physical Science, Hiroshima University, Higashi-Hiroshima, Hiroshima 739-8526, Japan} 
\altaffiltext{13}{ Solar-Terrestrial Environment Laboratory,  Nagoya University, Nagoya, Aichi 464-8602, Japan} 
\altaffiltext{14}{ National Astronomical Observatory of Japan, Mitaka, Tokyo 181-8588, Japan} 
\altaffiltext{15}{ School of Allied Health Sciences, Kitasato University, Sagamihara, Kanagawa 228-8555, Japan} 

\begin{abstract}
We observed the unidentified TeV gamma-ray source HESS J1804$-$216 with the CANGAROO-III atmospheric Cerenkov telescopes from May to July in 2006.
We detected very high energy gamma rays above 600 GeV at the $10\sigma$ level in an effective exposure of 76 hr.
We obtained a differential flux of $(5.0\pm 1.5_{\rm stat}\pm 1.6_{\rm sys}) \times 10^{-12}$(E/1\ TeV)$^{-\alpha}$ cm$^{-2}$s$^{-1}$TeV$^{-1}$ with a photon index $\alpha$ of $2.69 \pm 0.30_{\rm stat} \pm 0.34_{\rm sys}$, which is consistent with that of the H.E.S.S. observation in 2004. 
We also confirm the extended morphology of the source. 
By combining our result with multi-wavelength observations, we discuss the possible counterparts of HESS J1804$-$216 and the radiation mechanism based on leptonic and hadronic processes for a supernova remnant and a pulsar wind nebula.
\end{abstract}
\section{Introduction}
A Galactic plane survey was performed in 2004 by the H.E.S.S.\ 
imaging atmospheric Cerenkov telescope (IACT) with a flux sensitivity 
of 0.02~crab for 
gamma rays above 200\,GeV \citep{aha05a,aha06a}. Fourteen new
gamma-ray sources were detected at significance levels above 4$\sigma$, and
11 of the sources either have no counterpart or 
possible counterparts with significant positional offsets. 
HESS J1804$-$216 is one of the brightest, and its
spectrum is softest in this survey; the flux is about 0.25~crab above
200\,GeV with a photon index of $2.72\pm0.06$. In addition, with
a size of $\sim 22$~arcmin, it is one of the most extended TeV gamma-ray sources.  
The H.E.S.S.\ 
collaboration proposed two possible counterparts: the supernova remnant
(SNR) G8.7$-$0.1 and the young Vela-like pulsar B1800$-$21. However, 
the TeV gamma-ray source does not coincide exactly with either of these. 
G8.7$-$0.1 appears as a larger circular
region with a diameter of $\sim 50$~arcmin, and the geometric center
\citep{kas90b} has a large offset, $\sim11$ arcmin, from the centroid of
HESS J1804$-$216 to the northeast, while PSR B1800$-$21 has an
$\sim11$ arcmin offset to the west.

After the detection of HESS J1804$-$216, an SNR, G8.31$-$0.09, was discovered
at radio wavelengths \citep{bro06} to be located within the error circle of
HESS J1804$-$216, but with a smaller size, $5'\times4'$. {\it Suzaku} deep
observations discovered two new X-ray sources, Suzaku J1804$-$2142
(hereafter Src1) and Suzaku J1804$-$2140 (hereafter Src2), that are near the center
of HESS J1804$-$216 \citep{bam07}.  {\it Chandra} also detected these sources
\citep{kar07b}. {\it SWIFT} found three faint X-ray sources in the region of HESS
J1804$-$216 \citep{lan06}. One of them is positionally coincident with a
bright star, and another could also be associated with a star close to the
boundary of the XRT error circle. The other positionally coincides with
Suzaku Src2. So, there remain 5 possible counterparts: SNR G8.7$-$0.1, PSR
B1800$-$21, SNR G8.31$-$0.09, Suzaku Src1, and Suzaku Src2. We briefly
describe these sources in the following paragraphs.

SNR G8.7$-$0.1: 
G8.7$-$0.1 is associated with the W30 complex, which comprises
extended radio emission with a number of superposed smaller discrete
emission regions \citep{alt78, rei84, han87}. Radio recombination-line
observations have been used to identify discrete sources as H\,$\mathrm{II}$
regions, and CO observations also show molecular gas to be associated with
W30 \citep{bli82}. \citet{oje02} reported that massive star formation
may be occurring in molecular clouds in W30.
\citet{ode86} and \citet{kas90b} clearly established that G8.7$-$0.1 was an
SNR by the detection of a nonthermal extended radio emission. A {\it ROSAT}
observation revealed diffuse X-ray emission only from the northern half of
the remnant in the 0.1--2.4\,keV band \citep{fin94}. The distance to
G8.7$-$0.1 was estimated using several methods. Based on kinematical
distances to the H\,$\mathrm{II}$ regions associated with the SNR, the
distance was estimated to be $6 \pm 1$\,kpc \citep{kas90b}.
\citet{fin94} pointed out that more recent galactic rotation models applied
to the H\,$\mathrm{II}$ regions suggest a near kinematical distance of about
4.8\,kpc.  They also estimated the distance based on a Sedov solution
\citep{sed59, ham83} from the observed X-ray temperature and the angular
radius to derive 3.2\,kpc\,$\le d \le$\,4.3\,kpc for
an assumed initial energy of $10^{51}$ ergs. In this paper, we adopt $d =
4.8$\,kpc. They also estimated the age of the SNR, based on a Sedov solution
from the X-ray observation, to be 1.5--2.8\,$\times10^4$ years under the
assumption of an initial energy of $10^{51}$ ergs, and 2.7--3.9\,$\times10^4$
years under the assumption of a distance of 6\,kpc. Similarly, we estimated
an age of 2.2--3.1\,$\times10^4$ years under the assumption of a distance of
4.8\,kpc.
\citet{ode86} represented an age of $1.5\times10^4$ years from the relation
between the age and the surface brightness in the radio band.

PSR B1800$-$21:
The young Vela-like pulsar B1800$-$21 was found in a radio observation \citep{clif86}.
The offset from the centroid of HESS J1804$-$216 is $\sim$11~arcmin. 
The spin period, the spin period derivative, and the characteristic age are 
$P = 133.6$\,ms, $\dot{P}= 1.34\times10^{-13}$\,s/s, and 
$\tau _c=P/2\dot{P}= 15.8$\,kyears, respectively \citep{bri06}. 
The resulting spin-down luminosity is 
$\dot{E} = 4\pi ^2 I \dot{P}/P^3 = 2.2\times10^{36}(I / 10^{45})$
ergs\,s$^{-1}$, where $I$ is the moment of inertia in units of g\,cm$^2$. The
distance was estimated to be 3.9\,kpc from the pulsar's dispersion measure of
$233.99$\,pc\,cm$^{-3}$ \citep{tay93}. The newer Cordes \& Lazio NE2001
model \citep{cor02} gives a similar distance of 3.84$_{-0.45}^{+0.39}$\,kpc.
We adopt $d = 3.84$\,kpc throughout this paper. The association between PSR
B1800$-$21 and G8.7$-$0.1 has been discussed in several papers
\citep[e.g.][]{fin94,fra94,kas90a}.  However, a recent proper-motion
measurement \citep{bri06} has shown that the pulsar was born outside the
currently observed SNR, and that is moving more nearly toward the center of
G8.7$-$0.1, rather than away from it, which makes their association very
unlikely. Based on a 10\,ks observation with the {\it ROSAT} PSPC, Finley \&
\"{O}gelman (1994) reported a faint X-ray source near the radio pulsar
position, and attributed this emission to PSR B1800$-$21. Recently,
\citet{kar07a} and \citet{cui06} reported that an X-ray nebula around the
pulsar was detected with {\it Chandra}. Additionally, \citet{kar07a} reported that
the X-ray nebula has two structures: a brighter compact ($\sim 7'' \times
4''$) component (the inner pulsar wind nebula (PWN)) and an extended
($\sim12'' $) fainter emission component (the outer PWN). These are
asymmetric to the pulsar position and extended toward HESS J1804$-$216.

SNR G8.31$-$0.09: 
SNR G8.31$-$0.09 was found in a 90\,cm multi-configuration
Very Large Array survey of the Galactic plane. The size is $5'\times4'$, and
the offset from the centroid of HESS J1804$-$216 is 7~arcmin. The morphology
is shell-like and the spectral index is $\alpha_r = -0.6$ for $F_\nu \propto
\nu ^{\alpha_r}$ (Brogan et al.\ 2006).

Suzaku Src1: The offset from the centroid of HESS J1804$-$216 is 3~arcmin.
\citet{bam07} reported that Src1 is point-like or compact compared to
the spatial resolution of {\it Suzaku} with a half-power diameter of about 
2~arcmin. However, Kargaltsev et al.\ (2007b) reported that Src1 is extended or
multiple ($1.5'-2'$) with a {\it Chandra} observation. The {\it Suzaku} spectrum was
fitted with an absorbed power-law model \citep{bam07}. The best-fit
absorbing column is consistent with the Galactic hydrogen column in that
direction. Since the photon index of $-0.3 \pm0.5$ is very flat, \citet{bam07}
suggest that this source is likely to be a high-mass X-ray binary (HMXB). In
the {\it Chandra} observation \citep{kar07b}, no spectral fitting 
was able to be performed because of the low signal-to-noise ratio.  The unabsorbed
flux, which was estimated from the {\it Chandra} observation using the best-fit
parameters reported by \citet{bam07}, is a factor of $\approx$ 1.7 smaller
than that reported with the {\it Suzaku} observation of \citet{bam07}. The
difference could be due to unaccounted systematic errors, or the variability
of the source, which supports the HMXB interpretation \citep{kar07b}.
 
Suzaku Src2: The offset from the centroid of HESS J1804$-$216 is 1.8~arcmin.
Though \citet{bam07} reported that Src2 is extended or multiple, based on the
{\it Suzaku} observation, \citet{kar07b} reported that Src2 is point-like in
a {\it Chandra} observation. This could mean that the more sensitive {\it Suzaku}
observations have detected a fainter extended PWN component. The {\it Suzaku}
spectrum was fitted with an absorbed power-law model \citep{bam07},
and the best-fit absorbing column is about an order-of-magnitude higher than
the expected Galactic column.
This implies that Src2 is embedded in dense gas. The {\it Chandra} spectrum was
also well-fitted with an absorbed power-law model, and the obtained
absorbing column density is a factor of 2--3 larger than the Galactic column
\citep{kar07b}. \citet{kar07b} stated that the large absorption suggests
that Src2 is located within (or even beyond) the Galactic Bulge, or it shows
an intrinsic absorption, which is often seen in the X-ray spectra of HMXBs.
\citet{kar07b} also reported a marginal pulsation of 106~s in Src2, which
supports an HMXB interpretation. On the other hand, \citet{bam07} 
suggested that Src2 is a PWN or a shell-like SNR because of the extended
morphology observed with {\it Suzaku} and the best-fit photon index of 1.7
(0.7--3.1).

None of the above five sources morphologically match HESS
J1804$-$216. The counterpart is therefore still unknown.  In this paper, we
present TeV gamma-ray observations of HESS J1804$-$216 with the CANGAROO-III
telescopes and discuss the radiation mechanism and the counterpart by
combining our result with multi-wavelength observations.
\section{CANGAROO-III Observations}
CANGAROO-III is an array of four IACTs, 
located near Woomera, 
South Australia (136$^{\circ}47'$E, $31^{\circ}06'$S, 160\,m a.s.l.). 
Each telescope has a 10\,m diameter reflector 
made up of 114 segmented FRP spherical mirrors
mounted on a parabolic frame \citep{kaw01}. 
The telescopes are situated at the corners of a diamond 
with sides of $\sim$100\,m \citep{eno02}. 
The oldest telescope, T1, which was the CANGAROO-II telescope, 
was not used due to its smaller FOV and higher energy threshold. 
The imaging camera systems on the other three telescopes (T2, T3 and T4) 
are identical, with 427 PMTs and a FOV of 4.0$^{\circ}$ \citep{kab03}.
The PMT signals were recorded by charge ADCs and multi-hit TDCs \citep{kub01}.
The observations of HESS J1804$-$216 were made from May to July, 2006,
using the `wobble' mode in which the pointing position of each
telescope was shifted in declination by $\pm$0.5$^{\circ}$ from the centroid
of HESS J1804$-$216. The mean zenith angle of the observation was 21$^{\circ}$,
and the total observation time was 86.8 hr. We used the 3-fold
coincidence data taken at zenith angles of less than 40$^{\circ}$. To
trigger data recording, an individual telescope was required to have more
than four pixels receiving over 7.6 photoelectrons within 100\,ns (local
trigger), with a global trigger system to determine the coincidence of any
two of the three telescopes \citep{c3trig}.  We rejected data taken under
bad weather conditions in which the shower event rate was less than 6\,Hz.
%which is depending on the mean zenith angle and mirror reflectivity. 
Taking into account the DAQ dead-time, the effective live time was
calculated to be 76 hr.
%
%
%
%
%%%%%%%%%%%%%%%%%%%%%%%%%%%%%%%%%%%%%%%%%%%%%%%%%
%
%        analysis
%
%
%%%%%%%%%%%%%%%%%%%%%%%%%%%%%%%%%%%%%%%%%%%%%%%%%%%
\section{Data reduction and Analysis}
The basic analysis procedures are described in detail in \citet{eno06a} and
\citet{kab07}. Using calibration data taken daily with LEDs, the recorded
charges of each pixel in the camera were converted to the number of
photoelectrons. At this step we found 7 bad pixels out of 427 pixels for T2,
5 for T3, and 1 for T4, due to their higher or lower ADC conversion factors
in these observations. These bad pixels were removed from this analysis, which
was also reflected in the Monte Carlo simulations. After that, every shower
image was cleaned through the following CANGAROO-III standard criteria. Only
pixels that received $\ge$5.0 photoelectrons were used as ``hit pixels''.
Then, five or more adjacent hit pixels, with arrival times of within
30\,ns from the average hit time of all pixels, were recognized as a
shower cluster.

Before calculating image moments 
--- the ``Hillas parameters" \citep{hil85} ---
we applied the ``edge cut'' described in \citet{eno06b}. We rejected events
with any hits in the outer-most layer of the camera. The orientation angles
were determined by minimizing the sum of the squared widths with a
constraint given by the distance predicted by Monte Carlo
simulations.

We then applied the Fisher Discriminant method \citep{fis36,eno06a} 
with a multi-parameter set of $\vec{P} = (W_2,W_3,W_4,L_2,L_3,L_4)$, 
where $W$ and $L$ are the energy corrected width and length, 
and the suffixes represent the telescope IDs.
The Fisher Discriminant (FD) is defined 
as $FD \equiv \vec{\alpha}\cdot \vec{P}$, 
where $\vec{\alpha}$ is a set of coefficients mathematically determined 
in order to maximize the separation between two FDs for gamma rays and hadrons.

For a background study we selected a ring region around the target, $0.3\le
\theta ^2 \le0.5 $\,deg$^2$, where $\theta$ is the angular distance to the
centroid of HESS J1804$-$216 reported by the H.E.S.S.\ group \citep{aha06a},
and obtained the FD distributions 
for the background, $F_{b}$, and Monte Carlo gamma rays, $F_g$. 
Finally, we could fit the FD distributions of the events from the target
with a liner combination of these two components.
The observed FD distributions, $F$, should be represented as
$F = \alpha F_g + (1-\alpha)F_b$, 
where $\alpha$ is the ratio of gamma-ray events to the total number of events.
Here, only $\alpha$ was optimized. 
This analysis method was verified 
by an analysis of the Crab nebula data taken in December, 2005.

The reflectivities of each telescope, which were used in the Monte Carlo
simulations, were monitored every month by a muon ring analysis of a
calibration run taken individually by each telescope.  We obtained relative
light-collecting efficiencies with respect to the original mirror production
times of 0.60, 0.60 and 0.65 for T2, T3 and T4, respectively. Throughout
this analysis, we used the Monte Carlo simulations for gamma rays assuming a
point-source.
\section{Results}
\begin{figure}
\epsscale{}
\plotone{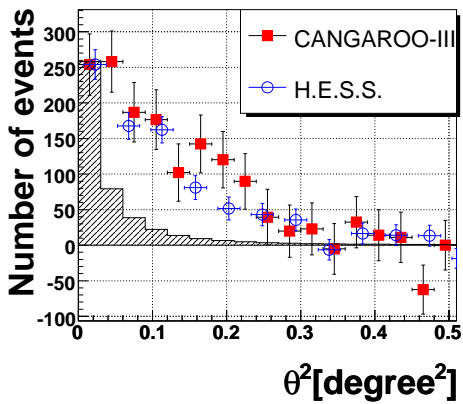}
\caption{
Number of excess events as a function of the squared angular distance. Here,
0$^{\circ}$ corresponds to the centroid of HESS J1804$-$216 reported by the
H.E.S.S.\ group \citep{aha06a}. The squares show the CANGAROO-III data
points. The circles show the normalized H.E.S.S.\ data points.  The hatched
histogram represents our PSF for a comparison.}
\label{theta}
\end{figure}

\begin{figure}
\epsscale{}
\plotone{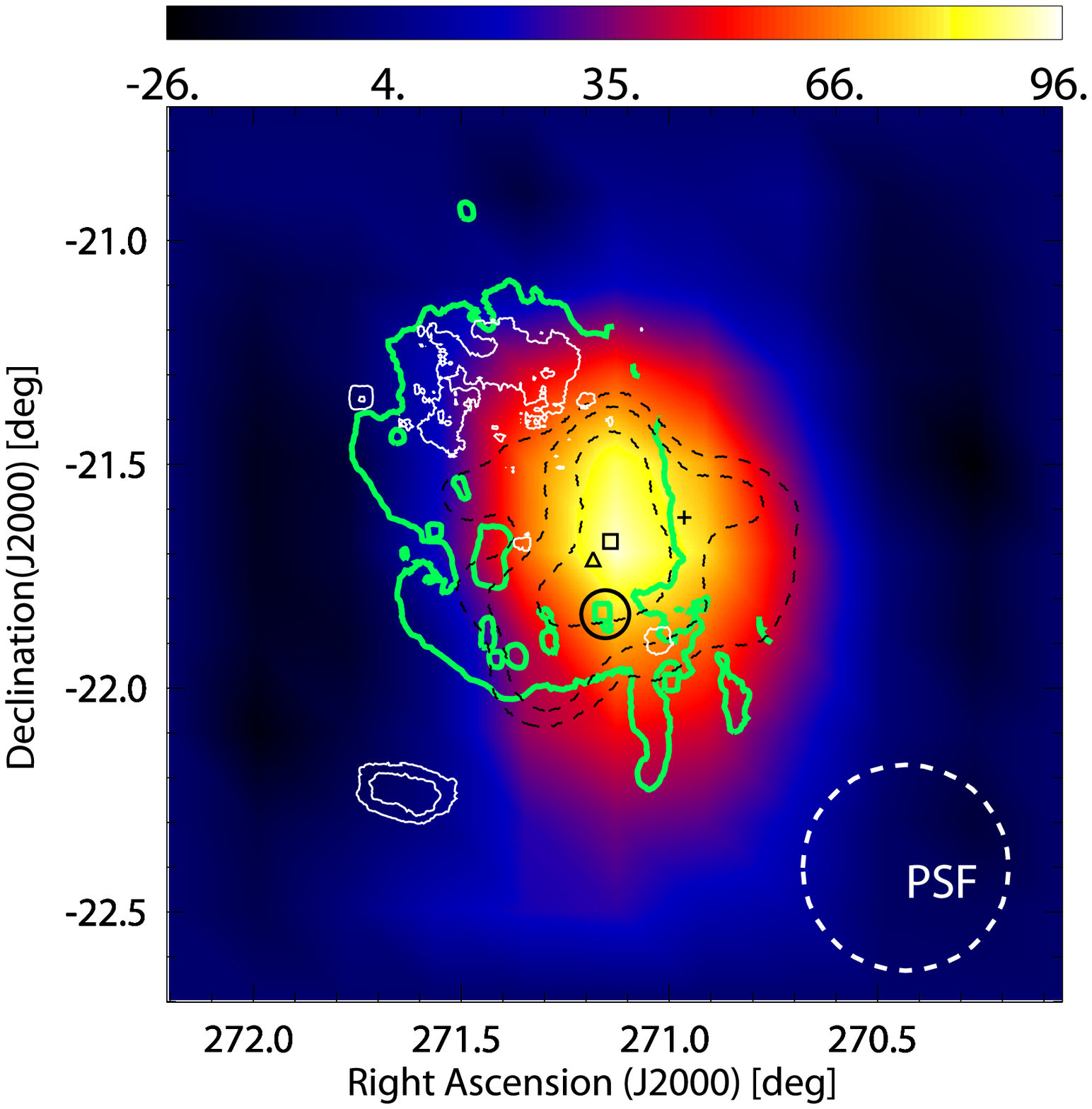}
\caption{
Smoothed morphology of gamma-ray--like events with our PSF of 0.23$^\circ$
radius.  Dashed contours show the VHE gamma-ray emission seen by H.E.S.S.\
 \citep{aha06a}. The thick solid contours (green) show the 20\,cm radio emission from
 G8.7$-$0.1 recorded by the VLA \citep{whi05}. The thin solid contours (white) show the
 X-ray emission detected by the {\it ROSAT} satellite \citep{fin94}. The solid circle indicates the position of G8.31$-$0.09 \citep{bro06}. The cross
indicates the PSR B1800$-$21 position \citep{bri06}. The triangle and
the square indicate the position of Suzaku Src1 and Suzaku Src2,
respectively.}
\label{map}
\end{figure}

The obtained $\theta ^2$ plot is shown in Fig.\,\ref{theta}
with the point spread function (PSF) of our telescopes, 
$0.23^{\circ}$ (68\% containment radius).
The numbers of excess events that we detected above 600\,GeV were $512 \pm61$
within $\theta ^2 <0.06$ deg$^2$, based on the assumption that it was a
point source, $977 \pm 94$ within $\theta ^2 <0.17$ deg$^2$, which
corresponds to that used in the spectral analysis by H.E.S.S.\ (and
taking into account the difference between our PSF and that of H.E.S.S.), and 
$1389 \pm 126$ within $\theta ^2 < 0.3$ deg$^2$.
The TeV gamma-ray emission is extended, and the morphology of gamma-ray--like
events, derived with boxcar smoothing with each pixel replaced by the average of its
square neighborhood, is shown in Fig.\,\ref{map}. The number of excess
events was individually estimated by the FD-fitting method in each
$0.2^\circ \times 0.2^\circ$ sky bin. When we evaluated the outer regions
($\theta ^2>0.6$ deg$^2$), we had to consider gradual deformations of the FD
distributions at larger angular distances from the target. Therefore, we
selected an annulus with radii $0.2^{\circ} < r < 0.4^{\circ}$ centered on the
evaluated region as the background. For the inner regions, $\theta ^2 <0.6$
deg$^2$, the events in $0.3\le \theta ^2 \le 0.5 $\,deg$^2$, excluding the
evaluated region, were adopted as a background.
The intrinsic extent of the TeV gamma-ray emission was estimated by a 2D
Gaussian fit on our unsmoothed excess map. The intrinsic deviations along
the Right Ascension and Declination axes were calculated to be
$0.160^\circ\pm 0.005^\circ$ and $0.274^\circ\pm 0.011^\circ$, respectively.
The best-fit centroid position was obtained (R.A, dec [J2000])$ =
$(271.079$^\circ$, -21.727$^\circ$).  The offset from the best-fit position
reported by H.E.S.S.\ \citep{aha06a} is ($\Delta$R.A, $\Delta$dec)$ =
$($-0.053^\circ\pm0.007^\circ$, $-0.026^\circ\pm 0.013^\circ$). The offset is not significant given our PSF.

\begin{figure} 
\epsscale{}
\plotone{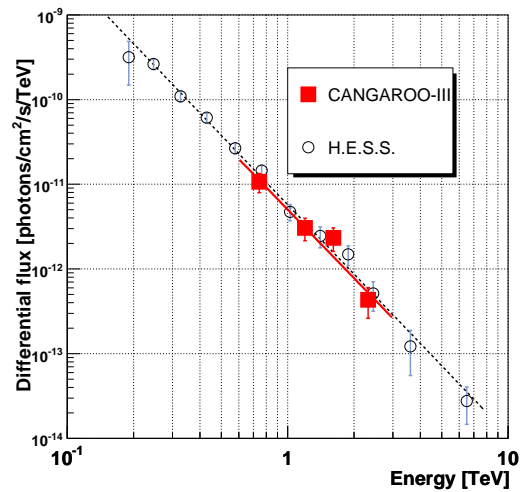}
\caption{Differential flux. The squares and circles show the CANGAROO-III
and the H.E.S.S.\ data points, respectively. The best-fit power-law is also
shown by the solid and dashed line from this work and from H.E.S.S., respectively.
}
\label{flux} 
\end{figure}

Figure \ref{flux} represents a reconstructed VHE gamma-ray differential spectrum
compatible with a single power-law: $(5.0\pm 1.5_{\rm stat}\pm 1.6_{\rm sys}) \times 10^{-12}$(E/1\ TeV)$^{-\alpha}$ cm$^{-2}$s$^{-1}$TeV$^{-1}$ with a photon index $\alpha$ of $2.69 \pm 0.30_{\rm stat} \pm 0.34_{\rm sys}$.
To obtain the spectrum, we used a cut of $\theta ^2 < 0.17$ deg$^2$. The
relevant systematic errors are due to the atmospheric transparency, night
sky background fluctuations, uniformity of camera pixels, and
light-collecting efficiencies.  
In addition, the signal integrating region
was changed from $\theta ^2 \le 0.17$ deg$^2$ to 0.3 deg$^2$, 
and the difference in fluxes was incorporated in the systematic errors.
The TeV gamma-ray extension and the flux obtained by
CANGAROO-III were consistent with those by H.E.S.S. Our result indicates
that the TeV gamma-ray emission was unchanged between the H.E.S.S.\
observations in 2004 and ours in 2006.
%\clearpage
%
\section{Discussion}
\begin{figure}[t]
\epsscale{}
\plotone{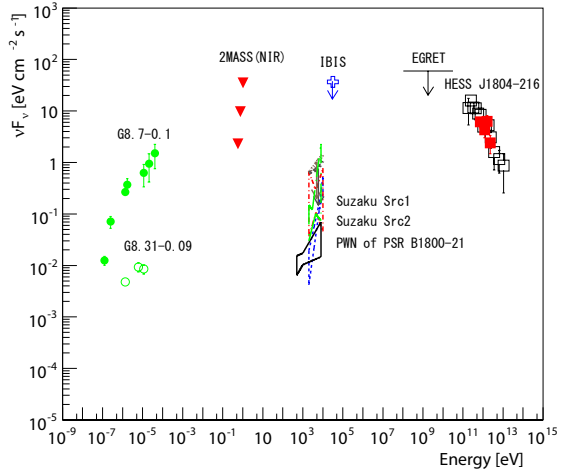}
\caption{Spectral energy distribution (SED) of HESS J1804$-$216, SNR G8.7$-$0.1,
  SNR G8.31$-$0.09, PWN of PSR B1800$-$21, Suzaku Src1, and Suzaku Src2 in all
  energy bands.  The data points derived from this work are represented by
  filled squares and references to others are given in Table~\ref{seddata}.
}
\label{Sed_allplot} 
\end{figure}

\begin{figure}[t]
\epsscale{}
\plotone{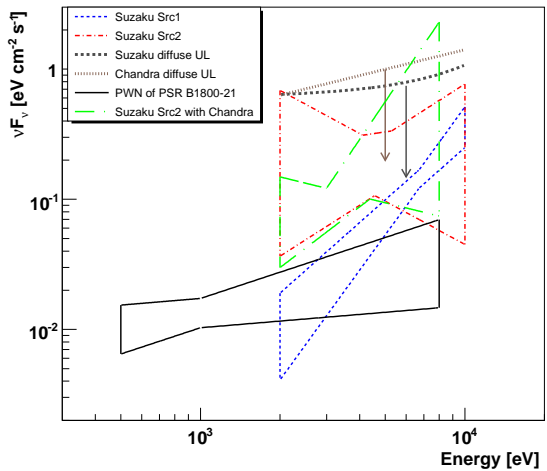}
  \caption{SED in the X-ray band. The dashed and dotted arrows show the
  upper limits for the diffuse source obtained by {\it Suzaku} and 
  {\it Chandra},
  respectively. The dashed, dot-and-short-dashed, dot-and-long-dashed, and
  solid closed regions show the error regions of Suzaku Src1, Suzaku Src2
  with {\it Suzaku} telescope, Suzaku Src2 with {\it Chandra} telescope, and PWN of PSR
  B1800$-$21 with {\it Chandra} telescope, respectively.
}
\label{Sed_xray} 
\end{figure}

As described in \S\,1, there are 5 possible counterparts: SNR G8.7$-$0.1,
PSR B1800$-$21, SNR G8.31$-$0.09, Suzaku Src1, and Suzaku Src2. 
Figures~\ref{Sed_allplot} and~\ref{Sed_xray} show the spectral energy
distribution (SED) of these counterparts and HESS J1804$-$216 including the
results of this work.  We extend the introductions to these sources given in
\S\,1 with a more detailed summary of the sources 
and their characteristics in the following paragraphs.

\citet{bam07} state that Suzaku Src1 and Src2 are physically associated
with HESS J1804$-$216. They suggested that X-rays and TeV gamma-ray emission
could come from an SNR shock, based on a model proposed by \citet{yam06}; in
an old SNR with an age of $\sim 10^5$ years, primary electrons have already
lost most of their energy, and only nucleonic cosmic rays remain.
Additionally, the old SNR shock colliding with a giant molecular cloud (GMC) 
can emit
hard nonthermal X-rays from secondary electrons and strong TeV gamma rays
from shock accelerated protons through $\pi^0$ decay. This model can explain
the observed large ratio of the TeV gamma-ray to X-ray flux (factors of
$\sim$100). The large X-ray absorption of Src2 also supports this scenario.

\citet{kar07b} suggested that a PWN is the source of HESS J1804$-$216, like
HESS J1825$-$137 or Vela~X. HESS J1825$-$137 is likely to be associated with 
the PWN G18.0$-$0.7 around the Vela-like pulsar B1823$-$13.  The TeV gamma-ray
emission detected with H.E.S.S.\ covers a much larger area than the X-ray
emission from G18.0$-$0.7, extending up to 1$^\circ$ southward from the
pulsar \citep{aha05b}. However, both the TeV gamma-ray and the
low surface-brightness X-ray emission have similarly asymmetric shapes, and
they are offset in the same direction with respect to the pulsar position. A
similar picture is observed around the Vela pulsar \citep{aha06b}. These
phenomena can be explained by the ``crushed PWN'' hypothesis \citep{blo01}: on
a time scale of $\sim 10^4$ years, the reverse SNR shock front propagates
toward the center of the remnant, where it crushes the PWN, and asymmetries in the
surrounding interstellar medium give rise to an asymmetric shape and offset
of the PWN relative to the pulsar and explosion site.

\citet{kar07b} considered the possibility that Suzaku Src1 or Src2 are a
PWN powering HESS J1804$-$216. However, the 3.24~s time resolution of the
{\it Chandra} ACIS observation precludes a search for the subsecond pulse
periods
expected for a young pulsar. Therefore, there is no strong evidence to
support it at this point.  They also suggested that PSR B1800$-$21
is associated with HESS J1804$-$216. Its asymmetric PWN component extending
toward HESS J1804$-$216, detected by {\it Chandra} \citep{kar07a}, shows a hint of
the association, but the sensitivity of the {\it Chandra} observation was possibly
insufficient to detect the PWN beyond 15$''$--20$''$ from the pulsar. 
Additionally, they pointed out that the extended morphology of HESS
J1804$-$216 argues against the HMXB interpretation because of the weak
observational evidence for HMXBs producing extended TeV gamma-ray
emission.

\citet{fat06} concluded that PSR B1800$-$21 cannot account for the spectrum
of HESS J1804$-$216, and G8.7$-$0.1 is probably the source of the TeV gamma
rays. However, they considered only a pion-decay model without any
consideration of the inverse Compton process, and for PSR B1800$-$21 they
only considered the acceleration of charged particles across voltage drops
in the relativistic winds near the light cylinder. Additionally, they did not
take into account the possibility of other sources besides PSR
B1800$-$21 and G8.7$-$0.1.

Based on the above discussions and the SED, we now discuss the radiation
mechanism of HESS J1804$-$216 and its counterpart.  For Suzaku Src1,
\citet{bam07} reported a very hard photon index of $-0.3\pm0.5$, which
corresponds to the power-law index of electrons of $-1.6\pm1.0$. 
This unusual value indicates that
the X-ray emission is not synchrotron radiation. Therefore, we do not discuss the SED
of Src1.
\citet{bam07} suggested that Src1 might be an HMXB. If it were, the
association between Src1 and HESS J1804$-$216 would be unlikely as discussed
in \citet{kar07b}.  The results of independent observations of the H.E.S.S.\
and CANGAROO-III telescopes show that HESS J1804$-$216 is quite extended
($\sim0.4^{\circ}$). This precludes the possibility that HESS J1804$-$216 is an
Active Galactic Nucleus. The plausible candidates seem to be an SNR or a
PWN.
%
%
%\clearpage

%\clearpage
%
%
%%%%%%%%%%%%%%%%%%%%%%%%%%%%%%%%%%%%%%%%%%%%%%%%%%%%%%%%%%%%%%%%%%%%%%
%
\subsection{SNR origin}
The SNR candidates associated with HESS J1804$-$216 are SNR G8.7$-$0.1, SNR
G8.31$-$0.09, and Suzaku Src2.

\ 

\begin{figure}[t]
\epsscale{}
\plotone{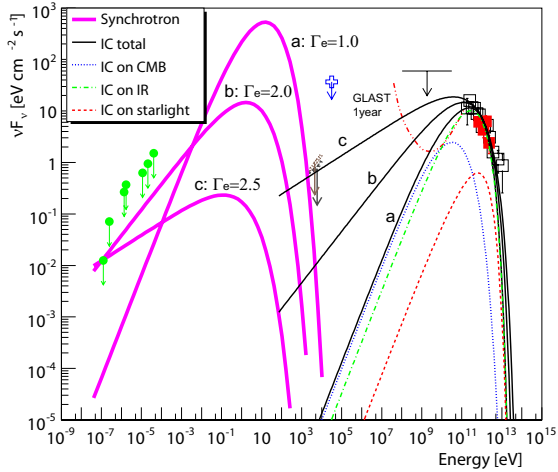}
\caption{Leptonic scenario for SNR G8.7$-$0.1.
  The thick solid curves show the synchrotron emissions, and the thin solid
curves show the total IC spectra with fixed spectral indices of $\Gamma_e =
1.0$ (a), $\Gamma_e = 2.0$ (b), and $\Gamma_e = 2.5$ (c), and with 
$B$ of 50\,$\mu$G, 8\,$\mu$G, 3\,$\mu$G, respectively.  The dotted, dot-dashed, and
dashed curves show the IC spectra (a) on each of CMB, IR and starlight,
shown in Fig.~\ref{isrf}, respectively. The dash-dot-dotted curve shows the
1 year, 5$\sigma$ sensitivity for the {\it GLAST} LAT taking into account the
diffuse background at the position of HESS J1804$-$216 \citep{ritz07}.
}
\label{Sed_diffuse} 
\end{figure}

\noindent
{\bf SNR G8.7$-$0.1:} 
We discuss the association between HESS J1804$-$216
and SNR G8.7$-$0.1. Figure~\ref{Sed_diffuse} shows the leptonic model curves
used to fit the TeV gamma-ray spectra obtained by H.E.S.S.\ and CANGAROO-III. 
We examined plausible inverse Compton (IC) model curves on the assumption
that the electron spectrum is a single power-law with an exponential cutoff,
\begin{equation}
\frac{dN_e}{dE_e} = K_eE _e
^{-\Gamma_e}\exp{(-E_e/E_{max\_e})} \ ,
\label{singlepl_el}
\end{equation}
where $K_e$ is the normalization factor, $E_e$ is the electron energy,
$\Gamma_e$ is the spectral index of the injected electrons,
and $E_{max\_e}$ is the maximum electron energy. In this paper, we do not
consider hard spectra of electrons with $\Gamma_e < 1$.

\begin{figure}[t]
\epsscale{}
\plotone{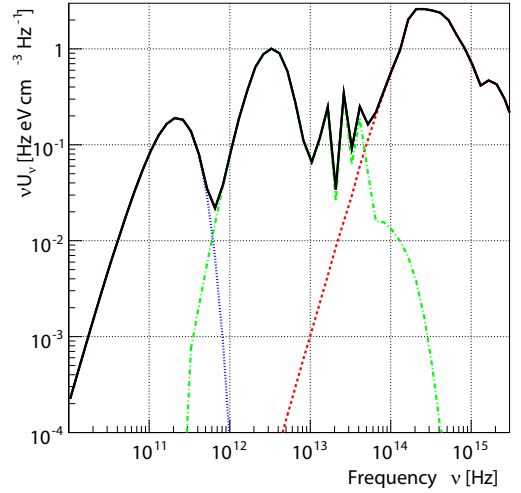}
\caption{
 Interstellar radiation field from the GALPROP package (v50p) at
 $(R,z)=(3.8, -0.01)$ kpc for G8.7$-$0.1.  From lower frequencies, CMB
 (dotted), IR emission from interstellar dust (dot-dashed), and optical
 photons from stars (dashed) are shown.  The solid line represents the sum
 of the three components.
 }
\label{isrf}
\end{figure}

To obtain the entire IC model curve, we used IR and optical (starlight)
photon fields for the target photons in addition to the cosmic microwave
background (CMB) field density. Here, we used an interstellar radiation
field (ISRF) derived from the latest (v50p) GALPROP package
\citep{por05,gal06}.
The ISRF was given for three components (CMB, IR from dust, and optical
starlight) as a function of the distance from the Galactic center, $R$ (in
kpc), and the distance from the Galactic plane, $z$ (in kpc). We extracted
the spectra at $(R,z)=(3.8, -0.01)$ at the SNR G8.7$-$0.1 position, as shown
in Fig.~\ref{isrf}. The X-ray upper limits for diffuse emission constrained
$\Gamma_e$ to be less than 2.5 for the IC model curves. We obtained
$E_{max\_e}$ to fit the IC model curves to the TeV gamma-ray spectra for
each fixed spectral index $\Gamma_e$ (Table~\ref{sum_ic}).

The electrons causing the IC scattering also emit synchrotron radiation. In
the radio band, the spectrum of the whole region of G8.7$-$0.1 is regarded
as the upper limits.  Here, the upper limits in the near-infrared (NIR) band,
in Fig.~\ref{Sed_allplot}, are neglected because they are for point sources. 
The magnetic field is constrained by the radio upper limits and the X-ray
upper limits for diffuse emission.  Table~\ref{sum_ic} lists the upper
limits of the magnetic field, maximum electron energy, and the total energy
of electrons above 0.51\ MeV,
\begin{equation}
W_e =\int^\infty _{0.51\ MeV} E_e\frac{dN_e}{dE_e}\ dE_e \ ,
\end{equation}
at $d=4.8$\,kpc for each electron spectral index $\Gamma_e$.
$W_e$ is estimated to be less than $\sim 10^{50}$ ergs for $\Gamma_e\le2.5$.
Although the total explosion energy of some supernovae are much higher, for
example, the total energy of SN2003lw, a type Ic supernova, is estimated to be
$\sim 6\times 10^{52}$\,ergs \citep{maz06}, the typical total energy of a
supernova explosion is estimated to be $\sim 10^{51}$\,ergs. If the
efficiency of the energy to accelerate the electrons is 10\%, the obtained $W_e$
can satisfy the SNR origin scenario.

\begin{figure}[t]
\epsscale{}
\plotone{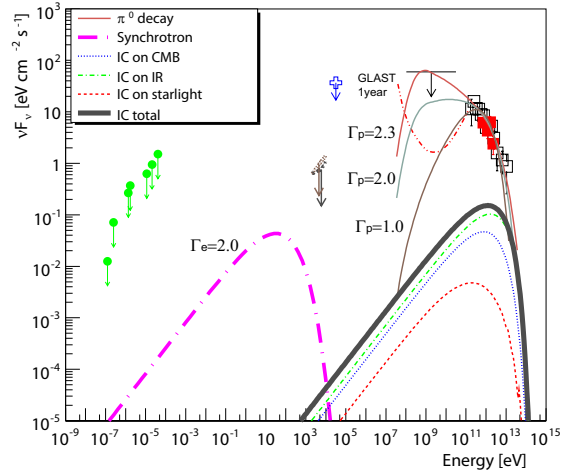}
\caption{Hadronic scenario for SNR G8.7$-$0.1.
The thin solid curves show the $\pi^0$ decay model curves with fixed
spectral indices of $\Gamma_p = 2.3$, $\Gamma_p = 2.0$, and $\Gamma_p =
1.0$, respectively. The thick solid curve shows the total IC spectra, and
the dot-and-long dashed curve shows the synchrotron emissions from the primary electrons with the
assumption of $\Gamma_e = 2.0$, $B=3$\,$\mu$G, and $K_{pe}=10^5(n/1$\ cm$^{-3})^{-1}$. The dotted, dot-dashed, and
dashed curves show the IC spectra on each of CMB, IR and starlight, shown
in Fig.~\ref{isrf}, respectively.
}
\label{Sed_pion} 
\end{figure}

Additionally, the hadronic scenario can also explain the TeV gamma-ray spectrum for G8.7$-$0.1.
In Fig.~\ref{Sed_pion}, we show the $\pi^0$ decay model curves with 
the assumption that the proton spectrum is a single power-law with an exponential cutoff,
\begin{equation}
\frac{dN_p}{dE_p} = K_p E_p
^{-\Gamma_p}\exp{(-E_p/E_{max\_p})} \ ,
\label{singlepl_pro}
\end{equation}
where $K_p$ is the normalization factor, $E_p$ is the proton energy,
$\Gamma_p$ is the spectral index of the injected protons,
and $E_{max\_p}$ is the maximum proton energy. In this paper, we do not
consider hard proton spectra with $\Gamma_p < 1$.  The EGRET upper
limit \citep{kar07b,hart99} constrained $\Gamma_p$ to be less than 2.3. We
obtained $E_{max\_p}$ to fit the $\pi^0$ decay model curves to the TeV gamma-ray spectra for each fixed spectral index $\Gamma_p$. 
Table~\ref{summary} lists the fitting parameters and the total energy of protons
above 1\,GeV,
\begin{equation}
W_p =\int^\infty _{1\ GeV} E_p\frac{dN_p}{dE_p}\ dE_p \ ,
\end{equation}
at a distance for SNR G8.7$-$0.1 of 4.8 kpc with the assumption that the
interstellar medium (ISM) density is $n = 1$ cm$^{-3}$.

The total energy of protons, $W_p$, is $9.3\times10^{51}(d/4.8 $
kpc)$^{2}(n/1$ cm$^{-3})^{-1}$\,ergs with $\Gamma_p = 2.3$. For a typical total
supernova explosion energy of $\sim 10^{51}$\,ergs,
if the efficiency of the energy to accelerate the protons is 10\%, the
ISM density should be $n\sim10^2$\,cm$^{-3}$. With the assumption of $\Gamma_p
=1.0 $, $W_p$ is $9.6\times10^{50}(d/4.8 $ kpc)$^{2}(n/1$ cm$^{-3})^{-1}$
ergs, and the ISM density should be $n\sim10$ cm$^{-3}$ for an efficiency of
10\%. However, \citet{fin94} derived the electron density $n_e$ to be
$(0.1-0.2)(d/6 $ kpc)$^{-1/2}$ cm$^{-3}$ for an X-ray emitting gas in the
remnant, based on the {\it ROSAT} observation.  In addition, the lack of the X-ray
emission in the TeV gamma-ray emission region, shown in Fig.~\ref{map},
indicates that the ambient density at the region is lower than that at the
X-ray emitting region, that is, $n < 0.1-0.2 $\,cm$^{-3}$. If it were, the
hadronic scenario would be unlikely.  On the other hand, a GMC, SRBY3, is near (in projection) to HESS J1804$-$216. The
offset, the angular radius, the distance, and the mass of the cloud are
$\sim 0.1^\circ$, $\sim 0.2^\circ$, 5.3 kpc, $55.6\times 10^4 M_{\odot}$,
respectively
\citep{cra02}. The mean number density is calculated to be $\sim 4.5 \times
10^2$\,cm$^{-3}$ under the assumption that it is spherical symmetry. If it
is associated with HESS J1804$-$216, there can be a high-density medium, and
the hadronic scenario can be satisfied.

Additionally, we estimated $K_{pe}$, the number ratio of protons to primary electrons. 
Under the assumption that the spectral indices of the protons and primary electrons are the same, $\Gamma_p = \Gamma_e$, $K_{pe}$ is derived from eqs.(\ref{singlepl_el})(\ref{singlepl_pro}) as $K_{pe} = K_{p}/K_{e}$.
Generally the maximum electron energy $E_{max\_e}$ is the same as $E_{max\_p}$, or $E_{max\_e}$ is lower than $E_{max\_p}$ due to a cooling effect, that is, $E_{max\_e} \le E_{max\_p}$. 
In Fig.~\ref{Sed_pion}, we show the leptonic model curves with the assumption of $K_{pe}\sim 10^5 (n/1$ cm$^{-3})^{-1}$, B\ =\ 3 $\mu$G, $\Gamma_e = 2.0$, and $E_{max\_e} = E_{max\_p} = 16$ TeV. The leptonic model curves move up with the lower $K_{pe}$.
With the assumption of B\ =\ 3$\mu$G, $K_{pe}$ is constrained to be 
$K_{pe} \ge 10^4 (n/1$ cm$^{-3})^{-1}$, because the IC model curve cannot exceed the TeV gamma-ray spectrum.
If a GMC is associated with HESS J1804$-$216, and $n\sim10^2$ cm$^{-3}$, the obtained $K_{pe}$ is consistent with that of the average cosmic rays in our Galaxy, $K_{pe}\sim 10^{2}$.
On the other hand, for a high $K_{pe}n$ of $\sim10^4$, we need to consider the contribution of the emission from secondary electrons produced by charged pions \citep{yam06,pfr04}. However, a detailed discussion about secondary electrons is beyond the scope of this paper.

\

\begin{figure}[t]
\epsscale{}
\plotone{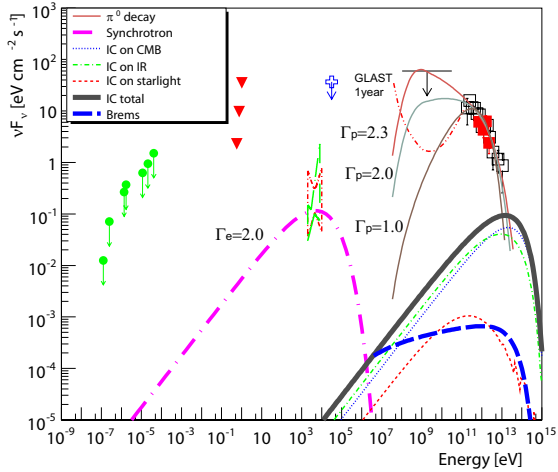}
\caption{SED with leptonic model curves to fit Suzaku Src2 and $\pi^0$ decay
model curves to fit TeV gamma rays. The thick solid curve shows the total IC spectra, and
the dot-and-long dashed curve shows the synchrotron emissions to fit Src2
with the assumption of a spectral index of $\Gamma_e = 2.0$ and
$B=3$\,$\mu$G. The dotted, dot-dashed, and dashed curves show the IC spectra
on each of CMB, IR and starlight, shown in Fig.~\ref{isrf}, respectively.
The long dashed curve shows the bremsstrahlung for the same electron spectrum
with the assumption of $n=1.0$\,cm$^{-3}$.  The thin solid curves show the
$\pi^0$ decay model curves with fixed spectral indices of $\Gamma_p =
2.3$, $\Gamma_p = 2.0$, and $\Gamma_p = 1.0$, respectively.
}
\label{Sed_point} 
\end{figure}

\noindent
{\bf Suzaku Src2:} For Suzaku Src2, we used the overlapping error region
obtained by {\it Suzaku} and {\it Chandra}.  In Fig.~\ref{Sed_point}, we examined
plausible synchrotron model curve within the error of the unabsorbed X-ray
flux on the same assumption as eq.~(\ref{singlepl_el}) with adopting 
$B$=3\,$\mu$G and $\Gamma_e = 2.0$.  The obtained IC model curves for the same
electron spectrum cannot reproduce the TeV gamma-ray spectrum.  Even if the
higher IR energy density than that in GALPROP is adopted, the cutoff energy of
the models does not change from $\sim10$ TeV, while the observed cutoff
energy should be $\le0.3$ TeV; such a low cutoff energy is due to the soft
spectrum of HESS J1804$-$216. A stronger magnetic field can reduce the
cutoff energy of the IC model curves.  In order to make the cutoff energy of
0.3\,TeV for Suzaku Src2, the magnetic field should be more than 7\,mG, and to
fit the TeV gamma-ray spectrum with the magnetic field the IR energy density
should be $\sim 10^6$ times higher than that in GALPROP. These values are
unlikely. We therefore conclude that the TeV gamma-ray spectra are not
produced by the IC model curves for the electron spectrum.

We also examined the bremsstrahlung emission for the same electron spectrum. To fit
the TeV gamma-ray spectrum, the ISM density should be more than 
$\sim 10^8$\,cm$^{-3}$, and the magnetic field should be stronger than 1.8\,mG for Suzaku
Src2. Due to such incredibly high values of the ISM density and the magnetic
field, we also reject the scenario that the bremsstrahlung 
radiation produces the TeV gamma-ray spectrum.

A simple solution to explain the observed TeV gamma-ray spectrum is that
accelerated protons produce the gamma rays, as shown in Fig.~\ref{Sed_point}. Therefore, the TeV gamma rays
and the X-rays could be produced by accelerated protons and electrons,
respectively. 
Assuming the spectral index, $\Gamma_p = \Gamma_e = 2.0$, the
maximum electron energy is $E_{max\_e} \le E_{max\_p} = 16$\,TeV 
(Table~\ref{summary}), and this maximum energy makes the lower limit of the
magnetic field of $B \ge 130$\,$\mu$G to fit the X-ray data. With the
magnetic field, we calculated the lower limit of
$K_{pe} \ge 6.6 \times 10^7 (n/1$ cm$^{-3})^{-1}$.
This value is much higher than that of the average cosmic rays in our
Galaxy, $K_{pe}\sim 10^{2}$.

\

\noindent
{\bf G8.31$-$0.09:} We consider the possibility that HESS J1804$-$216 is
associated with G8.31$-$0.09. For the leptonic scenario, it is difficult to
explain the fact that the size of G8.31$-$0.09 in the radio band is much smaller than
that in the TeV emission region.  On the other hand, if the accelerated
protons produced the TeV gamma rays, the difference in the size could be
explained due to the difference of the diffusion length between protons and
primary electrons.  However, we need additional information, either on the distance 
or from other wavelengths, to discuss the energetics or $K_{pe}$.
%\clearpage

%\clearpage

\subsection{PWN origin}
\begin{figure}[t]
\epsscale{}
\plotone{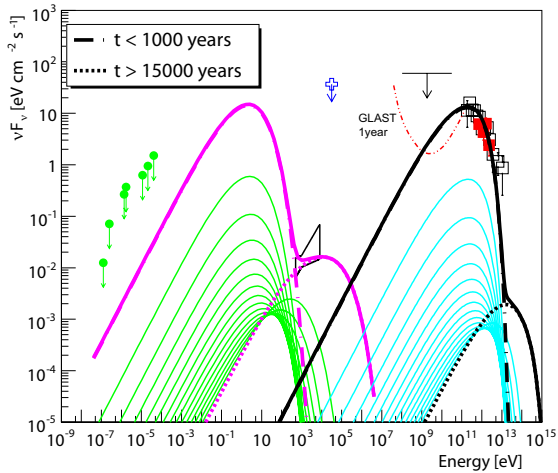}
\caption{
SED and the leptonic model curves for the PWN of PSR B1800$-$21 with a
time-dependent rate of electrons-injection with $\tau = 16$ kyears, $\tau_0
=700$ years, a braking index of $n_{br} \sim 1.6$, $\Gamma_e = 1.5$, and 
$B = 8$\,$\mu$G. The right-hand curves show the IC component, and the left-hand
curves show the synchrotron emissions. The dot-dashed curves show the spectra
produced by old electrons ($t < 1000$ years), and the dotted curves show the
spectra produced by young electrons ($t > 15000$ years). The thin solid
curves represent their medium per 1000 years ($1000< t < 15000$ years), and
the thick solid curves show their total.
}
\label{Sed_pwnall} 
\end{figure}

\begin{figure}[t]
\epsscale{}
\plotone{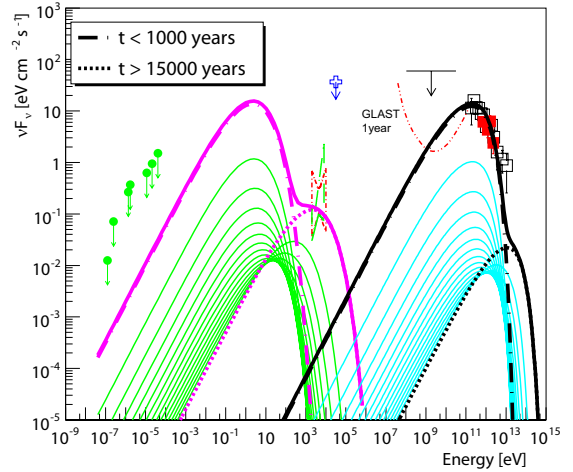}
\caption{
SED and the leptonic model curves for Suzaku Src2 with a time-dependent rate
of electrons-injection with $\tau = 16$ kyears, $\tau_0 =700$ years, a
braking index of $n_{br} \sim 1.8$, $\Gamma_e = 1.5$, and $B = 8$\,$\mu$G. The
model curves are the same as in Fig.~\ref{Sed_pwnall}.
}
\label{Sed_src2} 
\end{figure}
\begin{figure}[t]
\epsscale{}
\plotone{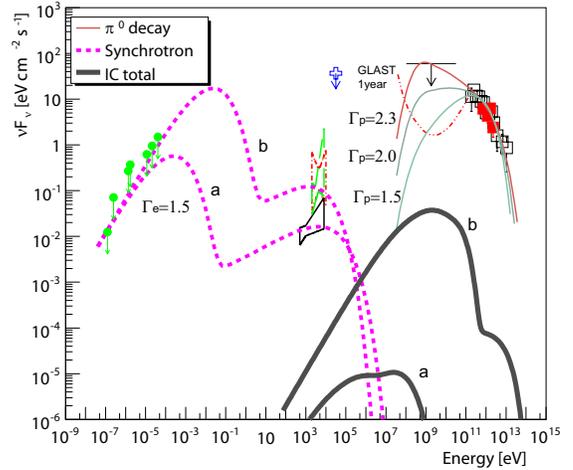}
\caption{
Hadronic scenario for PWN of PSR B1800$-$21\ (a) and Suzaku Src2\ (b). The
thin solid curves show the $\pi^0$ decay model curves with fixed
spectral indices of $\Gamma_p = 2.3$, $\Gamma_p = 2.0$, and $\Gamma_p =
1.5$, respectively. The thick solid curves show the total IC spectra, and
the dashed curves show the total synchrotron emissions from primary electrons with $\tau = 16$
kyears, $\tau_0 =700$ years, a braking index of $n_{br} \sim 1.6$, $\Gamma_e
= 1.5$, $B = 4$\,mG for PWN of PSR B1800$-$21~(a), and $B = 600$\,$\mu$G for
Src2~(b).
}
\label{Sed_hadronpwn} 
\end{figure}

At the present time, PWNe are the largest population of identified TeV
galactic sources, and the number of TeV gamma-ray sources that are located near a 
plausible pulsar candidate
are increasing \citep{hin07}. Therefore, it is very important to consider
the PWN origin. The PWN candidates associated with HESS J1804$-$216 are the
PWN of PSR B1800$-$21, and Suzaku Src2.

\

\noindent
{\bf PWN of PSR B1800$-$21:}
To estimate the total energy supplied from a pulsar, we should consider any
braking effects for spin-down luminosity.  If a pulsar spins down from an
initial spin period of $P_0$ according to $\dot{\Omega} = -k \Omega
^{n_{br}}$ assuming both $n_{br}$ and $k$ are constant, where $n_{br}$ is
the
braking index of the pulsar, the spin-down luminosity, $\dot{E}(t)$, is
given as
\begin{equation}
  \dot{E}(t) = \dot{E_0}\Bigl(1+\frac{t}{\tau _0}\Bigr)^{-\alpha}, 
  \>\> \alpha = \frac{n_{br}+1}{n_{br}-1} \>, 
\end{equation}
\begin{equation}
  \tau _0 \equiv \frac{P_0}{(n_{br}-1)\dot{P_0}} = \frac{P(t)}{(n_{br}-1)\dot{P}(t)} -t \>,
  \label{eqtau0}
\end{equation}
where $\tau_0$ is the initial spin-down timescale \citep{paci73}.
Therefore, the total energy that the pulsar has lost over its age, $\tau$, is 
\begin{equation}
E_{\rm pwn} =\int _0 ^{\tau}\dot{E}(t)dt= \frac{\dot{E_0}\tau _0}{1-\alpha}\Bigl[\Bigl(1+\frac{\tau}{\tau _0}\Bigr)^{1-\alpha} -1\Bigr].
\label{etot}
\end{equation}
For the Crab pulsar, produced in the supernova of 1054, 
a $\tau_0$ of $\sim 700$ years is derived
from eq.~(\ref{eqtau0}), and the total energy is derived as $E_{\rm pwn} =
3.8\times10^{49}$\,ergs with a braking index of $n_{br} = 2.5$ from
eq.~(\ref{etot}) \citep{liv06}.  For PSR B1800$-$21, the braking index and
the true age are unknown. The measured braking indices of other pulsars fall
in the range of $1.4\le n_{br} < 3$ \citep{liv06}. Therefore, we estimated
the total energies for several braking indices with $\dot{E}(\tau) =
2.2\times10^{36}$ ergs/s, an age of $\tau = 15.8$\,kyears (the same value as the
characteristic age), and $\tau_0 = 700$ years (the same as the Crab pulsar),
which are listed in Table~\ref{bindex}.

For the leptonic model, we assumed that the
rate of electrons-injection, $K_e'(t)$, proportionally decreases with
$\dot{E}(t)$, that is, $K_e'(t)\propto\dot{E}(t)$, and we considered the cooling effect due to
synchrotron and IC energy losses \citep{funk07}. The old electrons that were
injected with high-power initial spin-down luminosity have been cooled. On
the other hand the young electrons have less time to be cooled. 
Figure~\ref{Sed_pwnall} shows the SED with the leptonic model curves with a
time-dependent rate of electrons-injection with $\tau_0 = 700$\,years, a
braking index of $n_{br} \sim 1.6$, an age of $\tau = 16$\,kyears, $\Gamma_e
= 1.5$, and $B=$8\,$\mu$G: To obtain the model curves, we used the broken
energy due to synchrotron cooling and IC cooling, $E_{br}(t) \sim
3m_e^2c^3/(4\sigma_T(\tau-t)( B^2/(8\pi) +W_{ph} ))$, where $\sigma_T$ is
the Thomson cross-section and $W_{ph}$ is the total energy density of the
photon field. For the PWN of PSR B1800$-$21, we used the error region for
the entire PWN (inner + outer) obtained by {\it Chandra} \citep{kar07a}. The old
electrons are responsible for the TeV gamma-ray emission, and the young
electrons are responsible for the X-ray emission. This picture may also
explain the size difference between the TeV gamma-ray and the X-ray emission
regions, because the old electrons can extend further than the young ones.
To calculate the total energy of electrons, $W_e$, for the PWN origin scenario, 
we substituted $E_{br}(t)$ for $E_{max\_e}$.
Then, $W_e$ is estimated to be $\sim 1.7\times 10 ^{48}$\,ergs for the above
parameters. An $E_{\rm pwn}$ of $1.3\times10 ^{52}$\,ergs for the same parameters
can easily explain the amount of electron energy, and the efficiency of the
energy to accelerate electrons is $\sim1.3\times10^{-2}$\%. Note that the
above parameters are not unique, and other combinations of a $\tau_0$ and a
braking index are possible. If we adopt $\tau_0 = 30$ years, acceptable
model curves can be derived with a braking index of $n_{br} \sim 2.2$, an
age of $\tau = 16$ kyears, $\Gamma_e = 1.5$, and $B = 8$\,$\mu$G. For the latter parameters, $E_{\rm pwn}$ is estimated to be $2.3\times 10^{52}$\,ergs, and $W_e$
is $\sim 1.7\times 10^{48}$\,ergs.  The efficiency of the energy to
accelerate the electrons is $\sim7.4\times10^{-3}$\%.  In either case, a
quite high value for the total energy, $E_{\rm pwn}$, that the pulsar has lost
is required to explain the observed spectra.

We consider the hadronic model given in Fig.~\ref{Sed_hadronpwn}. Table~\ref{summary} lists the fitting parameters and the total energy of protons $W_p$ at a distance for PSR B1800$-$21 of 3.84 kpc with the assumption that the ISM density is $n = 1$ cm$^{-3}$. $W_p$ with $\Gamma_p$ of 2.3 is estimated to be
$5.9\times10^{51}(d/3.84 $ kpc)$^{2}(n/1$ cm$^{-3})^{-1}$\,ergs. With
a $\Gamma_p$ of 1.0, $W_p$ is estimated to be $6.1\times10^{50}(d/3.84 $
kpc)$^{2}(n/1$ cm$^{-3})^{-1}$\,ergs. The magnetic field is constrained to be
$B \ge 4$\,mG with the assumption of $\Gamma_p = 1.5$ and the constraint of
$E_{max\_e} \le E_{max\_p} = 7.9$\,TeV (Table~\ref{summary}). The magnetic
field is very high.
$K_{pe}$ is estimated to be more than $1.6 \times 10^4(n/1 $cm$^{-3})^{-1}$.
If a GMC is associated with HESS J1804$-$216, and $n\sim10^2$ cm$^{-3}$, the obtained $K_{pe}$ is consistent with that of the average cosmic rays in our Galaxy, $K_{pe}\sim 10^{2}$.
However, the lower limit of the magnetic field is much higher than the typical interstellar magnetic field, and there is no evidence supporting such a high magnetic field in this position.

\

\noindent
{\bf Suzaku Src2:}
For the leptonic model, a similar scenario as that for PWN of PSR B1800$-$21
can be considered for Suzaku Src2. However, because we have no information
about the age and the spin-down luminosity of Suzaku Src2, we assume the
same age and spin-down luminosity as that of PSR B1800$-$21. 
Figure~\ref{Sed_src2} shows leptonic model curves for Suzaku Src2 with $\tau_0 =
700$ years, a braking index of $n_{br} \sim 1.8$, an age of $\tau = 16$
kyears, $\Gamma_e = 1.5$, and B = 8 $\mu$G.
The total energy of the electrons is calculated to be $W_e = 1.8\times
10^{48}(d/3.84$ kpc)$^2$\,ergs, and $E_{\rm pwn}$ is $1.2\times 10^{51}$\,ergs.
This $E_{\rm pwn}$ would be proportionally reduced if $\dot{E}(\tau)$ is lower
than that of PSR B1800$-$21. For example, $E_{\rm pwn}$ is estimated to be $\sim
10^{50}$ ergs with the assumption of $\dot{E}(\tau) \sim 10^{35}$
ergs\,s$^{-1}$. 
Radio and X-ray observations with higher time resolutions might be able to
detect a sub-second pulse period from this source, and further observations in the
IR to UV band for a diffuse source in the extended region of HESS
J1804$-$216 could contribute to a tighter solution for the leptonic models.

We consider the hadronic model as given in Fig.~\ref{Sed_hadronpwn}. The
magnetic field is constrained to be $B \ge 600$\,$\mu$G with the assumption
of $\Gamma_p = 1.5$ and the constraint of 
$E_{max\_e} \le E_{max\_p} = 7.9$\,TeV (Table~\ref{summary}). 
$K_{pe}$ is estimated to be more than $1.5 \times 10^3(n/1 $cm$^{-3})^{-1}$. The lower limit of the magnetic field is much higher than the typical interstellar magnetic field.
\begin{deluxetable*}{lllc} 
\tablewidth{0pt}
%\rotate
\tablecaption{Summary of data used in the SED analysis.
\label{seddata}}
\tablehead{
  \colhead{Instrument}&
  \colhead{Source name}&
  \colhead{Symbol}&
  \colhead{Reference}
}
\startdata
VLA and other radio  & G8.7$-$0.1    & filled circle          & (1) \\
VLA                  & G8.31$-$0.09  & open circle            & (2) \\
{\it Suzaku} XIS     & Suzaku Src1   & closed region (dashed) & (3) \\
{\it Suzaku} XIS     & Suzaku Src2   & closed region (dot-and-short dashed)&(3) \\
{\it Suzaku} XIS     & diffuse UL (90\% C.L.)    & arrow (dashed)         & (3) \\
{\it Chandra} ACIS   & Suzaku Src2   & closed region (dot-and-long dashed)&(4) \\
{\it Chandra} ACIS   & PWN of PSR B1800$-$21 & closed region (solid)&(5) \\
{\it Chandra} ACIS   & diffuse UL    & arrow (dotted)         & (4) \\
{\it INTEGRAL} IBIS  & Upper limit   & arrow (cross)          & (4) \\
{\it EGRET}          & Upper limit   & arrow (solid)          & (4)(6) \\
{\it GLAST}          & 1 year sensitivity & dash-dot-dotted line & (7) \\
2MASS          & NIR point UL  & triangle               & (8) \\
H.E.S.S.       & HESS J1804$-$216 & open square         & (9) \\
CANGAROO-III   & HESS J1804$-$216 & filled square       & this work 
\enddata
\tablerefs{
  (1) Kassim \& Weiler 1990b;
  (2) Brogan et al.\ 2006;
  (3) Bamba et al.\ 2007;\\
  (4) Kargaltsev et al.\ 2007b;
  (5) Kargaltsev et al.\ 2007a;
  (6) Hartman et al.\ 1999;\\
  (7) {\it GLAST} LAT 2007;
  (8) Skrutskie et al.\ 2006;
  (9) Aharonian et al.\ 2006a
}
%\end{deluxetable}
\end{deluxetable*}
%%%%%%%%%%%%%%%%%%%%%%%%%%%%%%%%%%%%%%%%%%%%%%%%%%%
\begin{deluxetable}{cccc}
  \tablewidth{0pt}
  \tablecaption{Summary of parameters used in the leptonic model shown in Fig.~\ref{Sed_diffuse}.
    }
  \tablehead{
    \colhead{$\Gamma_e$} &
    \colhead{2.5} &
    \colhead{2.0} &
    \colhead{1.0}
  }
  \startdata
  $E_{max\_e}$ [TeV] & 3.6 & 2.3  & 1.3 \\
  $W_e$ [$10^{48}\rm ergs $] &330 &  17 & 1.4\\
  Upper limit of B [ $\mu$G ] & 1 & 8 & 50
  \enddata
\label{sum_ic}
\end{deluxetable}

\begin{deluxetable}{ccccc}
  \tablewidth{0pt}
\tablecaption{Summary of parameters used in the hadronic model.
\label{summary}}
\tablehead{
\colhead{$\Gamma_p$} &
\colhead{2.3} &
\colhead{2.0} &
\colhead{1.5} & 
\colhead{1.0} 
}
\startdata
$E_{max\_p}$ [TeV] & 30 & 16 & 7.9 &5.4\\
$W_p$\protect{\tablenotemark{a}}[$10^{51}\rm ergs$] & 9.3 &3.1 & 1.3 & 0.96\\
$W_p$\protect{\tablenotemark{b}}[$10^{51}\rm ergs$] & 5.9 &2.0 & 0.84 & 0.61
\enddata
\tablenotetext{a}{For SNR G8.7$-$0.1, $d=4.8$ kpc.}
\tablenotetext{b}{For PSR B1800$-$21, $d=3.84$ kpc.}
\end{deluxetable}
\begin{deluxetable}{ccc} 
  \tabletypesize{\scriptsize}
\tablewidth{0pt}
\tablecaption{Braking index dependence of the parameters.
\label{bindex}}
\tablehead{
  \colhead{braking index} &
  \colhead{$E_{\rm pwn}$ [ergs]} &
  \colhead{$P_0$ [ms]}
}
\startdata
  3& $2.6\times 10 ^{49}$ & 210\\
  2.5& $5.8\times 10 ^{49}$ & 120\\
  2.2& $1.3\times 10 ^{50}$ & 72\\
  1.8& $1.2\times 10 ^{51}$ & 19\\
  1.6& $1.3\times 10 ^{52}$ & 5.2\\
  1.4& $1.7\times 10 ^{54}$ & 0.37
  \enddata
\end{deluxetable}

\section{Conclusions}
CANGAROO-III observed HESS J1804$-$216, and detected gamma rays above 
600\,GeV at the $10\sigma$ level during an effective exposure of 76 hr. The
obtained differential flux is consistent with the previous H.E.S.S.\ result,
and the obtained morphology shows extended emission compared to our Point
Spread Function. We have discussed the radiation mechanism and considered
the proposed counterparts to the TeV gamma-ray source. 
For the SNR scenario, the most plausible counterpart 
%of HESS J1804$-$216 
is the SNR G8.7$-$0.1, and both hadronic and leptonic processes
can produce the observed TeV gamma-ray spectrum. We obtained upper limits of
spectral indices, $\Gamma_e \le 2.5$ for leptonic scenario, and $\Gamma_p
\le 2.3$ for hadronic scenario. The total energy of the primary 
electrons, $W_e$, is
estimated to be $3.3\times10^{50}(d/4.8 $ kpc)$^{2}$ ergs with $\Gamma_e =
2.5$. The total energy of the protons, $W_p$, is estimated to be
$9.3\times10^{51}(d/4.8 $ kpc)$^{2}(n/1$ cm$^{-3})^{-1}$\,ergs with $\Gamma_p
= 2.3$. The obtained $W_p$ indicates that a molecular cloud might be associated
to satisfy the hadronic scenario for the energetics. Extending this to include an estimation of the emission from secondary electrons will provide a tighter 
solution for hadronic scenario.
For the PWN scenario, we discussed both leptonic and hadronic processes. We confirmed that the leptonic model with a time-dependent rate of
electron-injection while considering the braking effect for the spin-down
luminosity and the cooling effect due to synchrotron and IC energy losses
could explain both the high TeV gamma-ray flux and the low X-ray flux.
However, a quite high value for the total energy that the pulsar has lost is
required for PSR B1800$-$21 in this model, $E_{\rm pwn} \sim 10^{52}$\,ergs. 
For Suzaku Src2, we could obtain an acceptable model curve, and $E_{\rm pwn}$ was
estimated to be $\sim 10^{50}$\,ergs with the assumption of $\dot{E}(\tau)
\sim 10^{35}$\,ergs\,s$^{-1}$. 
However, we need radio or X-rays observations to determine whether or not 
there is a pulse period in order to discuss the energetics further.

The model curves that are given in this paper indicate that {\it GLAST} could
determine the spectral index of the accelerated particles, and further
observations in the IR to UV band for a diffuse source could give a solution
of the counterpart and the radiation mechanism.

\acknowledgements
The authors would like to thank B.\ Lott for providing us the data of {\it GLAST} LAT sensitivity.
We also thank H.\ Yamaguchi for discussions on SNRs.
This work was supported by a Grant-in-Aid for Scientific Research by
the Japan Ministry of Education, Culture, Sports, Science and
Technology (MEXT), the Australian Research Council,
and the Inter-University Research Program of the
Institute for Cosmic Ray Research.
The work is also supported by
a Grant-in-Aid for the 21st century center of excellence programs
``Center for Diversity and Universality in Physics'' 
and ``Quantum Extreme Systems and their Symmetries''
from MEXT of Japan.
We thank the Defense Support Center
Woomera and BAE systems, and acknowledge all of the developers and
collaborators on the GALPROP project. 
Y.\ Higashi and T.\ Nakamori were supported by
Japan Society for the Promotion of Science Research Fellowships 
for Young Scientists.
%\clearpage
%

%\clearpage
%%%%%%%%%%%%%%%%%%%%%%%%%%%%%%%%%%%%%%%%%%%%%%%%%%%
%

%\clearpage
%\begin{deluxetable}{lllc} 
%\begin{deluxetable}{cccc} 
%\clearpage

\end{document}